\documentclass[%
 reprint,
 amsmath,amssymb,
 aps,
]{revtex4-2}

\usepackage{graphicx}
\usepackage{dcolumn}
\usepackage{bm}
\usepackage{mathrsfs}
\usepackage{float}      
\usepackage[table]{xcolor}      
\usepackage{subcaption} 
   
\begin{document}

\preprint{APS/123-QED}

\title{A general model of MnSi-like spiral magnets}

\author{K.~P.~W.~Hall}
\email{kpwhall@mun.ca}
\author{S.~H.~Curnoe}%
 \email{curnoe@mun.ca}
\affiliation{%
 Department of Physics and Physical Oceanography, Memorial University of Newfoundland,
St. John's, Newfoundland \& Labrador, Canada A1B 3X7
}%

\date{\today}

\begin{abstract}
A general, symmetry-allowed model of nearest-neighbour interactions for MnSi-like magnets is presented. A left-handed helical magnet phase occurs within a large parameter space of the model, which is explored via numerical simulation. The relations between microscopic features of the spiral structure and various model parameters, including an external magnetic field, are determined and show good agreement with predictions from free energy considerations. A skyrmion structure is stabilized near the boundary. 

\end{abstract}

                              
\maketitle

\section{Introduction}\label{sec:Intro}

The B20 structure type of magnetic crystals is renowned for displaying unusual magnetic structures and for applications in a number of different fields. These diatomic materials belong to the
non-centrosymmetric space group $\rm{P}2_{1}3$ 
which suports helimagnetic phases, arising from the absence of centrosymmetricity. 
As such, B20 materials have been at the forefront in the study of helical magnets where they were among the first to be experimentally observed \cite{williams_JApplPhys_1966, Lundgren_PhysScr.1_1970_Helical, IshikawaEtAl_SSC.19.1976}. In recent years there has been a resurgence of interest in these materials, MnSi in particular, due to the finding of non-Fermi liquid behaviour accompanied by partial magnetic order in the magnetic field-pressure-temperature phase diagram
\cite{pfleidererNature2001,doiron-leyraudNature2001,pfleidererNature2004, pfleidererScience2007,Tewari_2006PhysRevLett, Binz_2006PhysRevLett, Rossler2006_Nature,Fischer2008_PhysRevB, DoYiEtAl_PhysRevB.80.2009} and of a skyrmion crystal phase \cite{MuhlbauerEtAl_Science.323.2009}.

MnSi, an intermetallic compound, undergoes a 
phase transition at $T_c = 29$ K into a left-handed helical spin structure with wavevector ${\bf k}$ oriented along one of the $\langle111\rangle$ directions with a wavelength of 18 nm \cite{williams_JApplPhys_1966, Lundgren_PhysScr.1_1970_Helical, IshikawaEtAl_SSC.19.1976}. Under an applied magnetic field  greater than 100 mT ${\bf k}$ rotates to align 
with the field and a
conical spin structure is realized, where the cone 
angle decreases with field strength \cite{IshikawaEtAl_SSC.19.1976, LeBechJMMM1995_magnetic}. A field-induced ferromagnetic state appears above 
approximately 600 mT. A skyrmion crystal phase has been reported in `Phase A', a small pocket in the phase diagram just below $T_c$ for small magnetic fields ($\approx 100$ to 250 mT) \cite{MuhlbauerEtAl_Science.323.2009, Neubauer_2009_PhysRevLett}. A quantum phase transition occurs under pressure with $P_c = 14.6$ kbar accompanied by non-Fermi liquid behaviour over a wide range of temperature \cite{pfleidererNature2001,doiron-leyraudNature2001,pfleidererNature2004, menaPRB2004, pfleidererScience2007, Lee_2009PhysRevLett}.

As shown over sixty years ago, spiral magnetism arises from competing ferromagnetic (FM) and anti-ferromagnetic (AFM) interactions, which occur when there are more than one kind of exchange path \cite{Yoshimori_JPhysJapan.14.6.1959} or from anti-symmetric exchange interactions (known as the Dzyaloshinskii-Moriya (DM) interaction) that exist in non-centrosymmetric crystals
\cite{Moriya_PhysRev.120.1960,DzakoshinksiJPCS_1959}.
Over the years these phenomenological models of MnSi have been augmented by various additions, including gradient terms, same-site anistropy, generalized forms of the DM interaction,  and coupling to external fields, in order to quantitatively account for details of the helical structures (such as the wavenumber and orientation) and the nature of the phase transition to this state \cite{BakJensen_JPhysC.13.1980,PlumerWalker_JPhysC.14.31.1981, kataoka_JSPJ_1981, plumerJPhysC1984, HopkinsonKee_PRB.79.1.2009}.  The earliest studies employed a continuum approach to the magnetization which was later extended to lattice spin models for general three-dimensional crystals \cite{YiEtAl_PhysRevB.80.2009} and finally to the actual spin lattice for MnSi-type crystals \cite{HopkinsonKee_PRB.79.1.2009, ChizhikovDmitrienko_PhysRevB.85.2012}; the latter studies describe quantitatively not only the wave-number of the helical spin state, but also canting of the spins and field/pressure dependence.

Recently, muon spin rotation ($\mu$SR) studies have revealed more details of the magnetic structure in MnSi, including canting and rotation of spins within the spiral structure \cite{DalmasReotierEtAl_PhysRevB.93.2016,YaouancPhysRevResearch_2_2020}.
The main objective of this paper is to describe these details within a completely general model for MnSi-type crystals, constructed using only nearest-neighbour (NN) exchange interactions and single ion anisotropy.
This model will be analyzed via numerical simulations using the ``effective field method" (EFM) \cite{Walker_PRB.22_1980_Computer}. 


\section{The Model}\label{sec:Model}
The magnetic ions of B20 crystals occupy the $4a$ Wyckoff position of the space group P2$_{1}$3 ($T^4$, No.\ 198), forming a trillium lattice. 
The four spin sites within a cubic cell are: 
\begin{gather}\label{eq:Model:4aWyckoff}
\begin{aligned}
    \mathbf{r}_{1} = & \hspace{3em} (x,x,x ) \\
    \mathbf{r}_{2} = &\ ( -x + \frac{1}{2}, -x, x + \frac{1}{2} ) \\
    \mathbf{r}_{3} = &\ ( -x, x + \frac{1}{2}, -x + \frac{1}{2} ) \\
    \mathbf{r}_{4} = &\ ( x + \frac{1}{2}, -x + \frac{1}{2}, -x )
\end{aligned}
\end{gather}
where the parameter $x \approx 0.138$ for Mn in MnSi. In the following, a spin located at site ${\bf r}_i$ in any cubic cell will be called a \#$i$ spin. 

The underlying point group of $\rm{P}2_{1}3$ is the tetrahedral group $T$, which has  twelve symmetry elements. 
The corresponding operations in the space group include pure rotations and screw axes, but notably no inversion or reflections. None of the four sites of the $4a$ Wyckoff position are invariant under any space group operation.  

We will model magnetic interactions in B20 crystals by finding the most
general form of the nearest neighbour exchange interaction that is invariant under all space group operations. 
Each site has 
six nearest neighbours (NN) separated by a distance  $d =
a \sqrt{8x^2 - 4x +1/2}$ which gives
$d\approx .32a $, where $a = 4.558$ \AA\ is the lattice parameter. 
By considering all 
bilinears of the form $S_{i}^{\alpha} S_{j}^{\beta}$ (where $S_i^{\alpha}$ is the $\alpha$
component of the magnetic moment
at site $i$, and $j$ is a NN of $i$), it can be shown that these will combine to give $9$ invariants
consisting of $12$ bilinears each. These are
\begin{widetext}
\begin{gather}\label{eq:Model:Model}
\begin{aligned}
    H^{xx} & = \sum_{\mathbf{n}} S_{1\mathbf{n}}^{x} (S_{2\mathbf{n}}^{x} + S_{2\mathbf{n}'''}^{x})
    +(S_{3\mathbf{n}}^{x}+ S_{3\mathbf{n}'''}^{x}) S_{4\mathbf{n}'}^x
    +S_{1\mathbf{n}}^{z} (S_{3\mathbf{n}}^{z} + S_{3\mathbf{n}''}^{z}) \\
    & 
    +S_{2\mathbf{n'''}}^{z} (S_{4\mathbf{n}}^{z} + S_{4\mathbf{n}''}^{z})
    +S_{1\mathbf{n}}^{y} (S_{4\mathbf{n}}^{y} + S_{4\mathbf{n}'}^{y})
    +(S_{2\mathbf{n}}^{y}  +S_{2\mathbf{n'}}^{y}) S_{3\mathbf{n}''}^{y} \\
    H^{yy} & = \sum_{\mathbf{n}} S_{1}^{y} S_{2}^{y}+ S_{3}^{y} S_{4}^{y}+S_{1}^{x} S_{3}^{x}
    +S_{2}^{x} S_{4}^{x}
+S_{1}^{z} S_{4}^{z} + S_{2}^{z} S_{3}^{z}
  \\
    H^{zz} & = \sum_{\mathbf{n}} S_{1}^{z} S_{2}^{z}+ S_{3}^{z} S_{4}^{z} +
   S_{1}^{y} S_{3}^{y}+S_{2}^{y} S_{4}^{y}+  S_{1}^{x} S_{4}^{x}+S_{2}^{x} S_{3}^{x}\\
H_{s,a}^{xy} & = \sum_{\mathbf{n}} S_{1\mathbf{n}}^{x}( S_{2\mathbf{n}}^{y} \pm S_{2\mathbf{n'''}}^y) 
    + S_{1\mathbf{n}}^y( S_{2\mathbf{n'''}}^x
    \pm S_{2\mathbf{n}}^{x} )-
    (S_{3\mathbf{n}}^{x}\pm S_{3\mathbf{n}'''}^{x})
     S_{4\mathbf{n'}}^{y}
    - (S_{3\mathbf{n}'''}^y \pm S_{3\mathbf{n}}^y) S_{4\mathbf{n'}}^x \\
   & +  S_{1\mathbf{n}}^{z}( S_{3\mathbf{n}}^{x} \pm S_{3\mathbf{n''}}^x) 
    + S_{1\mathbf{n}}^x( S_{3\mathbf{n''}}^z
    \pm S_{3\mathbf{n}}^{z} )-
    S_{2\mathbf{n'''}}^{z}( S_{4\mathbf{n''}}^{x} \pm S_{4\mathbf{n}}^x) 
    - S_{2\mathbf{n'''}}^x( S_{4\mathbf{n}}^z
    \pm S_{4\mathbf{n''}}^{z} )
    \\
    & + S_{1\mathbf{n}}^{y}( S_{4\mathbf{n}}^{z} \pm S_{4\mathbf{n'}}^z) 
    + S_{1\mathbf{n}}^z( S_{4\mathbf{n'}}^y
    \pm S_{4\mathbf{n}}^{y} )
    -( S_{2\mathbf{n}}^y \pm S_{2\mathbf{n}'}^{y}  ) S_{3\mathbf{n''}}^z
    -    (S_{2\mathbf{n}'}^{z} \pm S_{2\mathbf{n}}^{z} ) S_{3\mathbf{n}''}^{y}  
    \\
    H_{s,a}^{yz} & = \sum_{\mathbf{n}} S_{1}^{y} S_{2}^{z}+ S_{3}^{y} S_{4}^{z}
    + S_{1}^{x} S_{3}^{y}+ S_{2}^{x} S_{4}^{y} 
    +S_{1}^{z} S_{4}^{x}+ S_{2}^{z} S_{3}^{x}  \\
    H_{s,a}^{zx} & = \sum_{\mathbf{n}} S_{1}^{z} S_{2}^{x}
    +  S_{3}^{z} S_{4}^{x}
    + S_{1}^{y} S_{3}^{z}+ S_{2}^{y} S_{4}^{z} 
    +S_{1}^{x} S_{4}^{y}+ S_{2}^{x} S_{3}^{y}
\end{aligned}
\end{gather}
\end{widetext}
where $\mathbf{n}$ is a cubic lattice vector, $\mathbf{n}' = \mathbf{n} -(1,0,0)$, $\mathbf{n}'' = \mathbf{n} -(0,1,0)$, and $\mathbf{n}''' = \mathbf{n} -(0,0,1)$. The terms in 
$H_{yy}$ and $H_{zz}$ have been abbreviated; their full forms are analogous to those in $H_{xx}$. Likewise the full forms of the terms in $H^{yz}_{s,a}$ and $H^{zy}_{s,a}$ can be constructed similarly to
$H^{xy}_{s,a}$. 
We also include the Zeeman term
\begin{equation}
\label{eq:Zeeman}
    H_{Z} = -\mathbf{H} \cdot \sum_{\mathbf{n},i} \mathbf{S}_{\mathbf{n},i}
\end{equation}
where $\mathbf{H}$ is an applied field. Lastly, we consider 
same-site anisotropy terms at second and fourth order:
\begin{gather}\label{eq:Model:Anisotropy}
\begin{aligned}
H_2 =& \sum_{\mathbf{n},i} \mathbf{S}_{i\mathbf{n}}^{x} \mathbf{S}_{i\mathbf{n}}^{y} + \mathbf{S}_{i\mathbf{n}}^{x} \mathbf{S}_{i\mathbf{n}}^{z} + \mathbf{S}_{i\mathbf{n}}^{y} \mathbf{S}_{i\mathbf{n}}^{z}, \\
H_4 = & \sum_{\mathbf{n},i} |\mathbf{S}_{i\mathbf{n}}|^4.
\end{aligned}
\end{gather}

 Computational studies typically use fewer parameters 
by taking specific combinations of
the terms given in
(\ref{eq:Model:Model}).
For example, the NN Heisenberg exchange interaction is
\begin{equation}
    \sum_{\langle i,j \rangle} \mathbf{S}_{i} \cdot \mathbf{S}_{j} =  H^{xx} +  H^{yy} + H^{zz}.
\end{equation}
The NN Dzyaloshinski-Moriya (DM) interaction is
\begin{equation}
    \sum_{\langle i,j \rangle} \mathbf{D}_{ij} \cdot  \left( \mathbf{S}_{i} \times \mathbf{S}_{j} \right),
\end{equation}
where the DM vectors $\mathbf{D}_{ij}$ are constrained by the symmetry of the lattice. This interaction corresponds to the three antisymmetric (subscript $a$) terms in (\ref{eq:Model:Model}); in fact comparison with these terms yields the minimal constraints on the DM vectors. For example, examining the terms in $H_{a}^{xy}$ yields
$D_{1\mathbf{n}2\mathbf{n}}^z = -D_{1\mathbf{n}2\mathbf{n}'''}^z = -D_{3\mathbf{n}4\mathbf{n}'}^z = D_{3\mathbf{n}'''4\mathbf{n}'}^z = -D_{1\mathbf{n}3\mathbf{n}}^y = D_{1\mathbf{n}3\mathbf{n}''}^y \ldots 
$
The DM vectors are often further constrained by taking the DM term to be $H^{xy}_a + H^{yz}_a + H^{zx}_a$. Also, the symmetric terms $H^{\alpha\beta}_s$ are usually omitted altogether. 

In our simulations, we consider a free energy constructed using all nine of the exchange terms (\ref{eq:Model:Model}) and the two same-site anisotropy terms (\ref{eq:Model:Anisotropy}), for a total of eleven independent interaction constants, as well as the Zeeman term (\ref{eq:Zeeman}):
\begin{eqnarray}
  F &= &  {\cal J}^{xx}H^{xx} +  {\cal J}^{yy}H^{yy} +
    {\cal J}^{zz}H^{zz} + {\cal J}^{xy}_sH^{xy}_s 
    +  {\cal J}^{xy}_aH^{xy}_a  \nonumber \\
    & & + \ldots +  
    {\cal J}_2 H_2 + {\cal J}_4 H_4 + H_Z .
\end{eqnarray}


\section{Magnetic Order Parameters}\label{sec:OP}

The helical magnet phase of MnSi is marked by the appearance of a left-handed spiral oriented along one of the four equivalent 
$\langle 111 \rangle$ directions. From here on, we will assume that the helices are oriented in the particular direction $[111]$. Perpendicular to 
$[111]$ are alternating planes of 
\#1 spins and of \#2, 3 and 4 spins, which we 
label 1 and 2-3-4 respectively. 
Moving along the $[111]$ direction, the distance between a 1-plane and the following 2-3-4-plane is $(1-4x)a/\sqrt{3} = 0.259 a$,
and the distance between a 2-3-4-plane and the next 1-plane is $4x a/\sqrt{3} = 0.319 a$.

In order to classify magnetic structures
we define magnetic order parameters
 associated with  $\mathbf{k}~\lvert\rvert~[111]$.
The little group  of this wavevector
is $C_{3}$, a subgroup of the crystallographic point group $T$. The 
three representations that belong to  $\mathbf{k}~\lvert\rvert~[111]$, 
labelled  $F_1$, $F_2$ and $F_3$, are derived from the three
representations of $C_3$. All three are one-dimensional; $F_2$ and $F_3$ are related by time reversal.

The $12$-dimensional basis of $4$ spins within a unit cell and $3$  spatial dimensions generates a (reducible) representation belonging to $\mathbf{k}~\lvert\rvert~[111]$,
$4F_1 \oplus 4 F_2 \oplus 4 F_3$ -- that is, there are four copies of each irreducible representation. 
The basis is:
\begin{eqnarray}\label{eq:op1}
F_{1}^{(1)} & = & S^{x}_{1\mathbf{k}} + S^{y}_{1\mathbf{k}} + S^{z}_{1\mathbf{k}}\\
F_{2}^{(1)} & = & S^{x}_{1\mathbf{k}} + \varepsilon S^{y}_{1\mathbf{k}} + \varepsilon^2 S^{z}_{1\mathbf{k}}\\
F_{3}^{(1)} & = & S^{x}_{1\mathbf{k}} + \varepsilon^2 S^{y}_{1\mathbf{k}} + \varepsilon S^{z}_{1\mathbf{k}}\\
F_{1}^{(2)} & = & S^{x}_{2\mathbf{k}} + S^{y}_{4\mathbf{k}} + S^{z}_{3\mathbf{k}}\\
F_{2}^{(2)} & = & S^{x}_{2\mathbf{k}} + \varepsilon S^{y}_{4\mathbf{k}} + \varepsilon^2
S^{z}_{3\mathbf{k}}\\
F_{3}^{(2)} & = & S^{x}_{2\mathbf{k}} + \varepsilon^2 S^{y}_{4\mathbf{k}} + \varepsilon
S^{z}_{3\mathbf{k}}\\
F_{1}^{(3)} & = & S^{y}_{2\mathbf{k}} + S^{z}_{4\mathbf{k}} + S^{x}_{3\mathbf{k}}\\
F_{2}^{(3)} & = & S^{y}_{2\mathbf{k}} + \varepsilon S^{z}_{4\mathbf{k}} + \varepsilon^2 S^{x}_{3\mathbf{k}}\\
F_{3}^{(3)} & = & S^{y}_{2\mathbf{k}} + \varepsilon^2 S^{z}_{4\mathbf{k}} + \varepsilon S^{x}_{3\mathbf{k}}\\
F_{1}^{(4)} & = & S^{z}_{2\mathbf{k}} + S^{x}_{4\mathbf{k}} + S^{y}_{3\mathbf{k}}\\
F_{2}^{(4)} & = & S^{z}_{2\mathbf{k}} + \varepsilon S^{x}_{4\mathbf{k}} + \varepsilon^2 S^{y}_{3\mathbf{k}}\\\label{eq:op12}
F_{3}^{(4)} & = & S^{z}_{2\mathbf{k}} + \varepsilon^2 S^{x}_{4\mathbf{k}} + \varepsilon S^{y}_{3\mathbf{k}},
\end{eqnarray}
where $\varepsilon = \exp 4\pi i /3 = -\frac{1}{2} - i \frac{\sqrt 3}{2}$ and $\varepsilon^2 = \varepsilon^{*}$.   $S^{\alpha}_{i\mathbf{k}}$ are the Fourier transforms of the spins,
$$
S^{\alpha}_{i\mathbf{k}} = \frac{1}{N^{1/2}} 
\sum_{\mathbf{n}} \exp(-i \mathbf{k}\cdot \mathbf{r}_{i\mathbf{n}}) S_{i\mathbf{n}}^{\alpha}
$$
where  $\mathbf{r}_{\mathbf{n}}$ is
  the $\mathbf{n}$th lattice vector and $N$ is the total number of cells. The physical spins are 
  $S_{i\mathbf{n}}^{\alpha}$, that is
  \begin{equation}
    \frac{1}{N^{1/2}} \sum_{\bf k} S_{i\mathbf{k}}^{\alpha} \exp(i \mathbf{k}\cdot \mathbf{r}_{i\mathbf{n}}).
  \end{equation}
  There is no sum over wavenumber in this expression when ${\mathbf k}$ is the only wavevector present in the structure.

The helical magnet phase of MnSi is marked by the appearance of a left-handed spiral, which corresponds to a order parameter that transforms as $F_3$ {\em i.e.}, one or more of the $F_3^{(i)}$ are non-zero while 
$F_2^{(i)} = 0$. That is, the spins precess in a clockwise direction with respect to the $[111]$ direction. Since $F_1$ is compatible with $F_3$ (in the sense that no additional symmetries are broken by $F_1$) there is no requirement that $F_1^{(i)}$ be vanishing.

The spin arrangements associated with a $F_3$ order parameter are in general
quite complicated; here we make some simplifying assumptions based on experimental observations. First, we assume that the magnitude of individual spins is fixed. 
If $F_3^{(1)}$ is present and not $F_2^{(1)}$ then this constraint forces $F_1^{(1)}$ to be absent in order for spin \#1 to have fixed length. This means that spin \#1  must be perpendicular to the $[111]$ axis - {\em i.e.}, no canting of this spin
toward $[111]$ is expected, as seen in experiment.

If we also assume that spins \#2, 3, and 4
have the same magnitudes then the relative magnitudes of the various order parameters will be constrained, but there is no requirement that $F_1^{(i)}$ must vanish if  $F_3^{(i)}$ is present and not $F_2^{(i)}$  for $i=2,3,4$.  However, in experiments it is observed that spins \#2, 3, and 4 lie perpendicular to ${\bf k}$ in ferromagnetic arrangements within each 2-3-4-plane \cite{DalmasReotierEtAl_PhysRevB.93.2016}. In this arrangement, if $F_2^{(i)}$ vanishes then $F_1^{(i)}$ also vanishes and the components of $F_3^{(i)}$ are related by
$F^{(2)}_3 = \varepsilon^2 F^{(3)}_3 = \varepsilon F_3^{(4)}$. 

Even with all these constraints imposed there remains a free parameter accessible by experiment: the relative orientation of spins in a 1-plane with respect to the nearest 2-3-4-plane \cite{DalmasReotierEtAl_PhysRevB.93.2016}. The spins precess in a left-handed sense for a distance $d$ along the $[111]$ direction by an angle $k d$ rad, 
where $k = 0.35$ nm$^{-1}$. The precession angle between planes of the same type is therefore $ka/{\sqrt{3}}$ rad or  5.28$^{\circ}$. According to this simple picture, the precession angle between a 2-3-4-plane and the next 1-plane would be 2.913$^{\circ}$; however this angle is measured to be only 0.86$^{\circ}$ \cite{DalmasReotierEtAl_PhysRevB.93.2016}. The phase difference $\phi = -2.04^{\circ}$ (-0.0356 rad)
can be considered as a model-dependent parameter.

The only constraint we enforce in our numerical simulations is that all four spins have constant, equal magnitudes. We find that spins in the 2-3-4-planes do not always lie perpendicular to ${\bf k}$. Generally, they precess within a different plane, which gives rise to another model-dependent parameter, the angle $\gamma$, which we define as the angle between the $(111)$ plane and the plane of the of the spins precession. Since there are two kinds of planes, there are two angles, $\gamma^{(1)}$, for the 1-plane and $\gamma^{(2)}$, for the 2-3-4-plane. These angles have not been measured in experiment, but they have been predicted in quantitative analysis~\cite{ChizhikovDmitrienko_PhysRevB.85.2012}. Furthermore, the individual spins are not ferromagnetically aligned, instead each \#$i$ spin will cant towards the axis $\tau_i$,
where
\begin{gather}\label{eq:CantDir}
\begin{aligned}
    \mathbf{\tau}_{1} & = [111]  \\
    \mathbf{\tau}_{2} & =
    [\bar{1}1\bar{1}]  \\
    \mathbf{\tau}_{3} & =
    [\bar{1}\bar{1}1] \\
    \mathbf{\tau}_{4} & =
    [1\bar{1}\bar{1} ].
\end{aligned}
\end{gather}

\section{Results}\label{sec:Result}
The EFM is a computational method used for determining the
spin configuration of a system as $T \rightarrow 0$ by finding local minima of the free energy in classical and semi-classical systems with pairwise interactions.
The method uses an iterative algorithm which, in each step, scans all spins in a random order. For each spin site, a \textit{local field} is calculated,
\begin{equation}
    H_{i}^{\alpha} = - \sum_{j,\beta} \mathscr{J}_{i,j}^{\alpha,\beta} S_{j}^{\beta}
\end{equation}
where  ${\bf S}_{j}$ are the spins with which ${\bf S}_{i}$  interacts and $\mathscr{J}_{i,j}^{\alpha,\beta}$ is the total of all interaction constants for $ S_{i}^{\alpha}$ and $ S_{j}^{\beta}$. 
The spin located at $i$ is then reoriented, either fully or partially, in the direction of this field. 
This process is repeated many times until a local minimum is found. Since the algorithm has no process by which a given spin can increase its interaction energy, it is likely that the final lattice configuration will not be the global minimum. This is remedied by running the algorithm a large number of times with randomly generated starting configurations. From this set of simulations, only those which produce the lowest energy are selected.

Since we are primarily interested in modelling the 
helical phase, we begin by considering a set of interaction constants which yields such a configuration, and then we vary those parameters in order to discern their individual effect on the magnetic structures.
For each simulation, we measure
 the wavenumber $k$, the phase difference $\phi$, the out-of-plane canting angle $\gamma$ (with respect to the $[111]$ plane) and the relative size of the magnetic order parameters 
$|F_i^{(j)}|$. All simulations were performed on a system with $23 \times 23 \times 23$ cells (large enough to contain one full wavelength of the helix)
with 48668 spins in total. In order to find an incommensurate $k$, periodic boundary conditions were {\em not} imposed.

\subsection{Reduced Model}\label{ssec:Results:DMLike}
\graphicspath{ {./Sections/5_Results/img/} }

We begin by examining a model with only two parameters, $J$ and $D$, defined by
\begin{eqnarray}
J & = & {\cal J}^{xx} =  {\cal J}^{yy} = {\cal J}^{zz} \\
D &=& {\cal J}^{xy}_a =  {\cal J}^{yz}_a = {\cal J}^{zx}_a,
\end{eqnarray}
corresponding to the Heisenberg exchange interaction and a simplified DM interaction where all components of the DM vector are $\pm D$. In all simulations, we used ${\cal J}_2 =-J/2$, ${\cal J}_4 = J/2$  with all other constants zero. The sign of $D$ is negative, yielding a 
left-handed spiral. 

\begin{figure}[H]
\centering
\includegraphics[width=0.45 \textwidth, keepaspectratio]{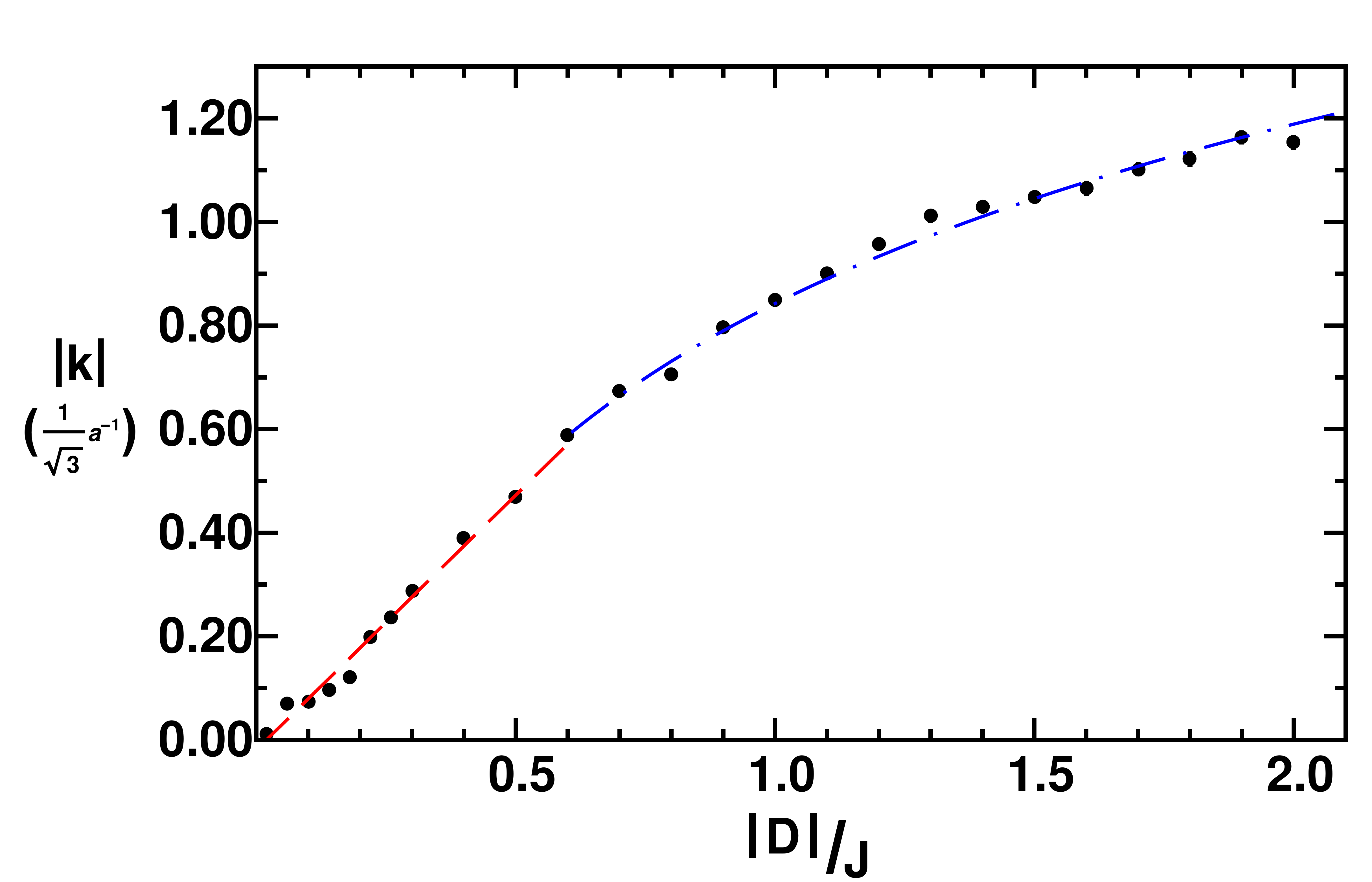}
\caption[]{Average wavevector  $k$ as a function of $|D|/J$.
For $|D|/J <  0.6$ a linear fit is shown (red; dashed). For $|D|/J > 0.6$ a logarithmic fit is shown (blue; dot-dashed).}
\label{fig:kvsD}
\end{figure}

The wavelength was measured using the average rotation between unit cells along $[111]$.
As shown in Fig.\ \ref{fig:kvsD}, for small $|D|$, we find a linear
relationship between wavenumber $k$ and $|D|$, as predicted from free energy considerations \cite{BakJensen_JPhysC.13.1980}. Non-linear deviations occur for $|D|/J \gtrapprox 0.6$.

The measured wavelength is
$\lambda = 18$ nm $= 22.8 \sqrt{3} a$, that is  22.8 unit cells along the $\langle111\rangle$ direction. The corresponding wavenumber is 
$k = 0.276/(\sqrt{3} a)$, which occurs for $|D|/J \approx 0.3$.

\begin{figure}[H]
\centering
\includegraphics[width=0.45 \textwidth, keepaspectratio]{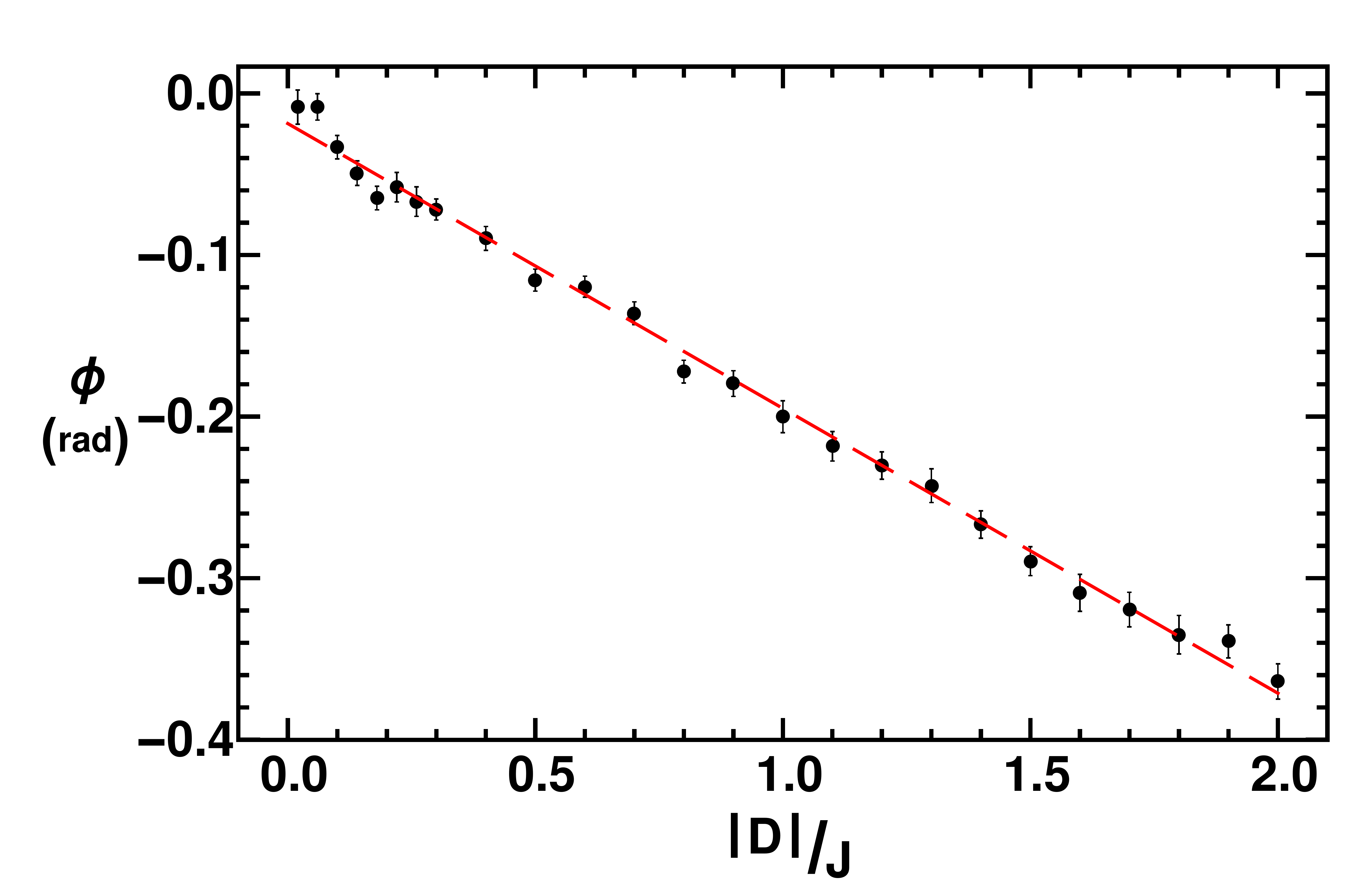}
\caption[]{Phase difference $\phi$ as a function of $|D|/J$.  A linear fit is shown.}
\label{fig:phivsD}
\end{figure}

Fig.\ \ref{fig:phivsD} shows the phase difference
$\phi$  (defined at the end of Section III) as a function 
of $|D|/J$. $\phi$ is the average of phase differences measured between a 1-plane and a 2-3-4-plane. For small $|D|/J$, there is a  linear relation between $\phi$ and $|D|/J$, as predicted from free energy considerations \cite{ChizhikovDmitrienko_PhysRevB.85.2012}.

The value of $|D|/J \approx 0.3$ yields a phase difference
$\phi \approx -0.07$ rad, which is approximately twice as large as the measured value \cite{DalmasReotierEtAl_PhysRevB.93.2016}.

\begin{figure}[H]
\centering
\includegraphics[width=0.45 \textwidth, keepaspectratio]{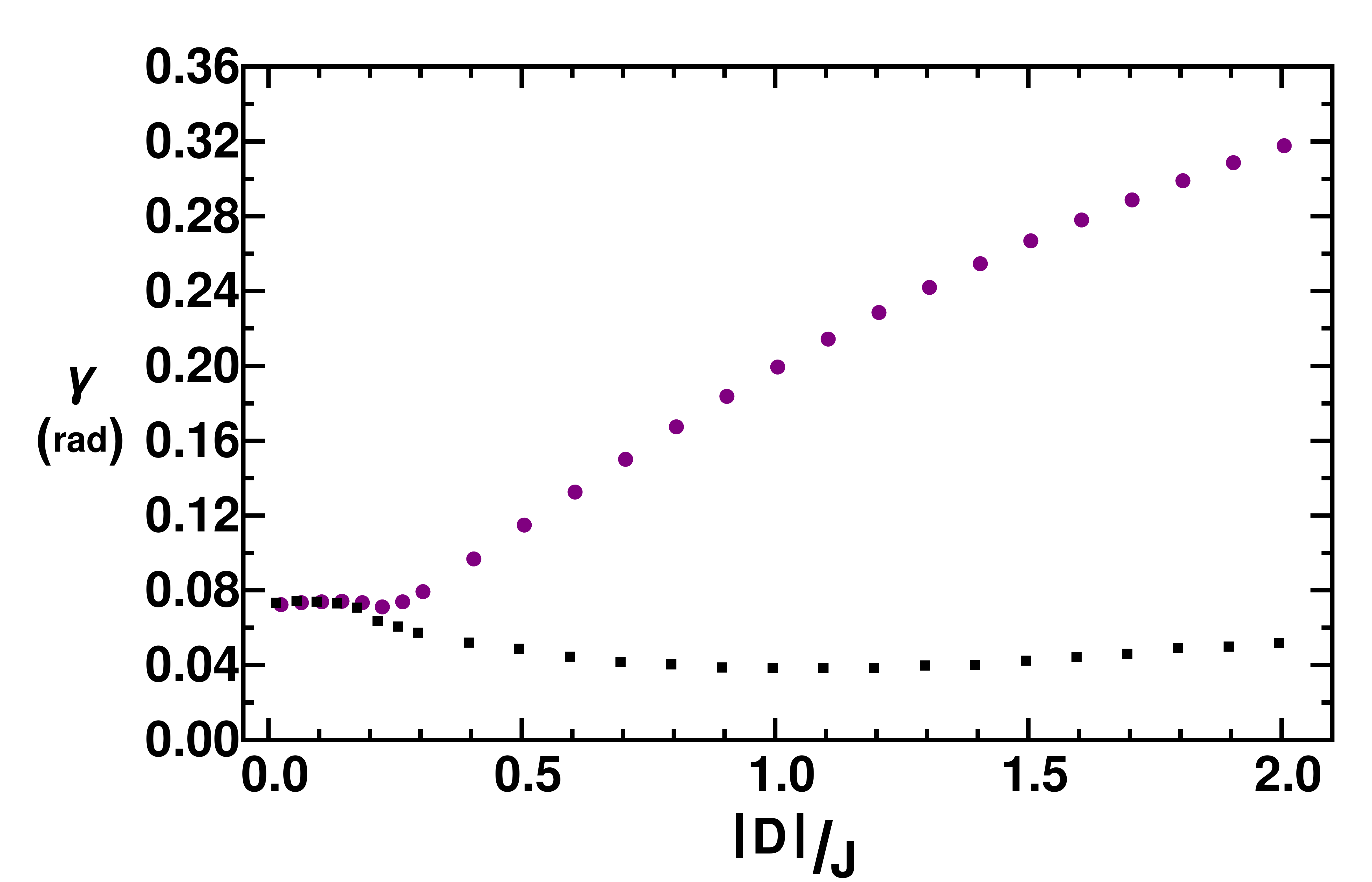}
\caption[]{The average out-of-plane angle $\gamma^{(1)}$ for the 1-plane (black squares) and 
$\gamma^{(2)}$ for the 2-3-4-plane (purple circles). In the 2-3-4-planes, the individual spins in a given layer have the same orientation.}
\label{fig:gammavsD}
\end{figure}

\begin{figure}[H]
\centering
\includegraphics[width=0.5 \textwidth, keepaspectratio]{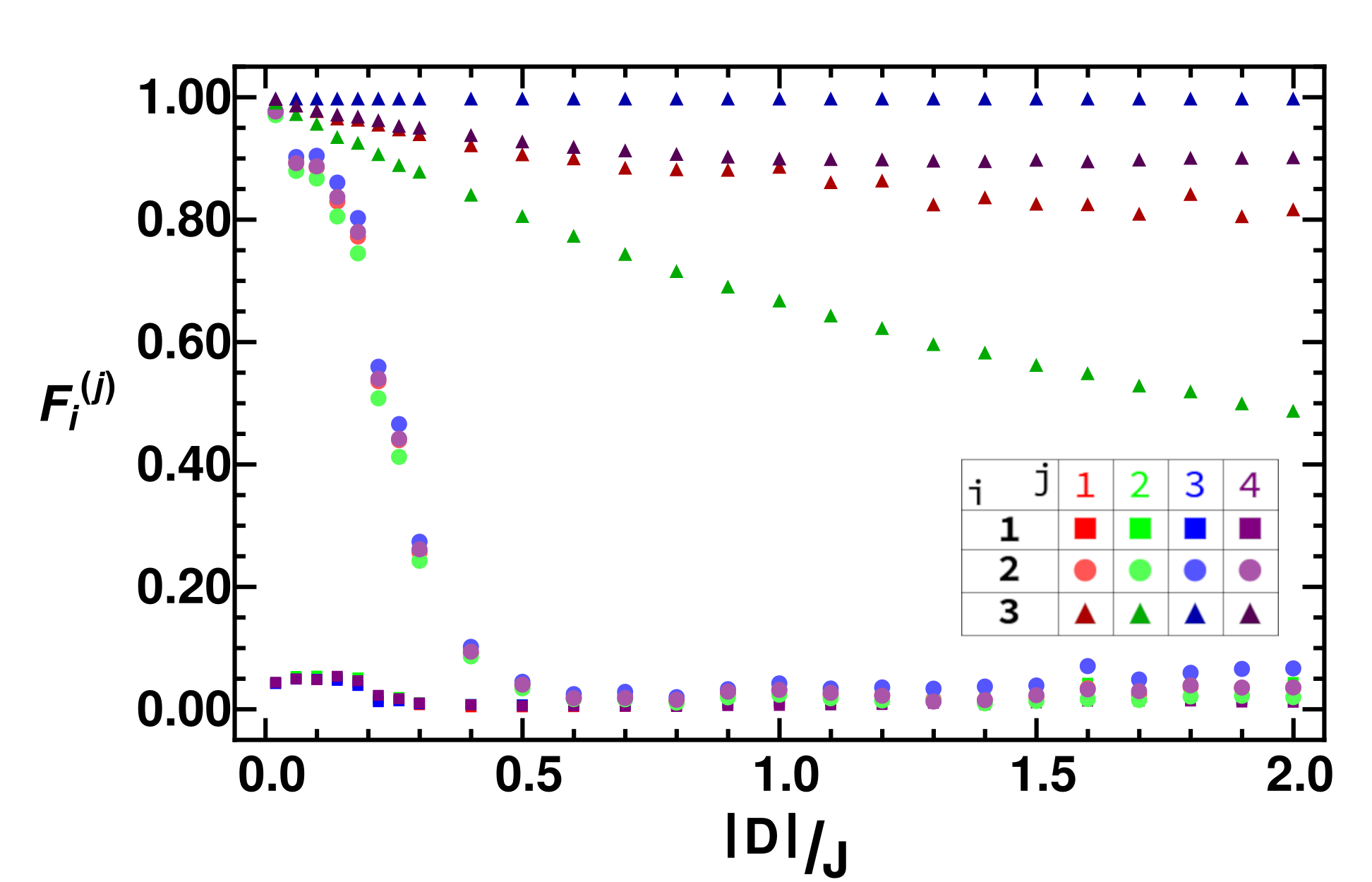}
\caption[]{The averaged magnetic order parameters derived from  Eqs.\ 8-19 as a function of $|D|/J$.  The plots are normalized 
such that the maximum size of any order parameter at each value of $|D|/J$ is always $1$.}
\label{fig:OP}
\end{figure}

Fig.\ \ref{fig:gammavsD} shows the  angle
$\gamma$ (defined at the end of Section III) for each kind of plane as a function 
of $|D|/J$. In our simulations, $\gamma$ is measured by assuming that all spins associated with a given position rotate within the same plane and measuring the angle between the $[111]$ axis and the normal vector of this plane. $\gamma$ is the average of angles measured for
each spin. The canting of spins in a 1-plane is always small, which correlates with a vanishing $F_1$ order parameter, as shown in Fig.\ \ref{fig:OP}. However, the canting of the  spins in a 2-3-4-plane increases with $|D|/J$, as expected from the free energy considerations \cite{ChizhikovDmitrienko_PhysRevB.85.2012}. Using the value $|D|/J \approx 0.3$ (determined from our measurements of  $\mathbf{k}$), 
we estimate that  $\gamma \approx 4.6^{\circ}$ for the 2-3-4-plane in MnSi. 

Fig.\ 4 shows magnetic order parameters, calculated by taking the absolute value of the functions $F_i^{(j)}$ defined in Eqs.~\ref{eq:op1}--\ref{eq:op12}, as a function of $\lvert D \rvert$/J. The summations are calculated using the assumption that $\mathbf{k}~\lvert\rvert~[111]$, with the magnitude of $k$ determined by our simulations (see Fig.~1).

In order to make it easier to compare the relative sizes of the order parameters, the plots in Fig.\ 4 have been normalized at each value of $|D|/J$ such that the highest value is one. In fact, the scale of each plot
decreases with $|D|/J$, indicating a transfer of weight to other values of $k$ with increasing $|D|/J$. This is most likely due to finite size and non-periodic boundary conditions of our simulation.

Fig.\ 4 shows that at larger wavelengths 
(small $|D|/J$) there is a mixture of $F_2$ and $F_3$ order parameters, corresponding to
right-handed and left-handed structures, but the
right-handed part quickly decreases with respect to the left-handed part as the wavelength decreases. At
$|D|/J \approx 0.3$, the left-handed component is five times larger than the right-handed component, and all four parts of the left-handed order parameter are present, but not equal.


There is a slight increase of $F_1$ order parameters with larger values of $|D|/J$, which coincides with the appearance of isolated skyrmions in our simulations, such as the one shown in Fig.\ 5. When present, skyrmions always appear near the boundary as a small tunnel through a few layers.

\begin{figure}[H]
    \centering
    \includegraphics[width=0.45 \textwidth]{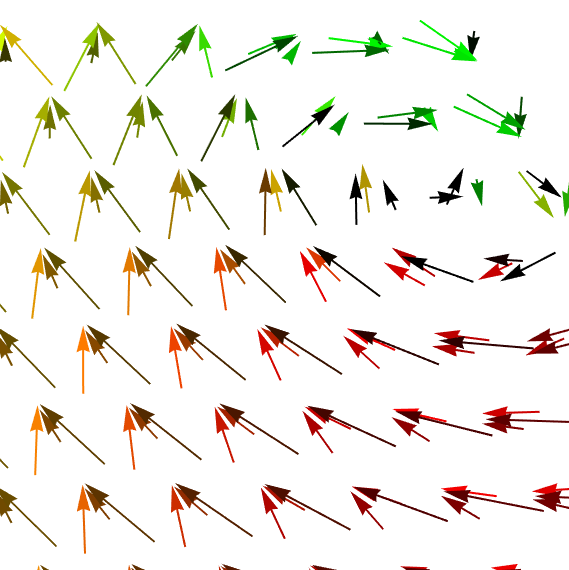}
    \caption{Example of a skyrmion spin configuration for $|D|/J = 2.00$. The hue of the arrow varies with the angle within the $(111)$ plane, while the brightness represents the magnitude of the out-of-plane angle (darker arrows have a larger out-of-plane angle).}
\end{figure}

\subsection{Individual model terms}\label{ssec:Results:Individual}
In this Section, we examine the effects of each independent interaction constant. As the starting point, we use a set of parameters 
which yield a left-handed helix: $J=1$, $D= -0.5$,  ${\cal J}_2 =-1/2$, ${\cal J}_4 = 1/2$, with all other constants vanishing. We vary the nine
interaction constants around this point in order to examine the 
dependence of $k$, $\phi$ and $\gamma$ on these parameters. In the following, we present those
results where the dependence on individual 
parameters is most pronounced. Complete details of all of the simulations may be found in Ref. \cite{hall_thesis}.

\begin{figure}[H]
    \begin{subfigure}[c]{0.45 \textwidth}
        \includegraphics[width=\textwidth,keepaspectratio]{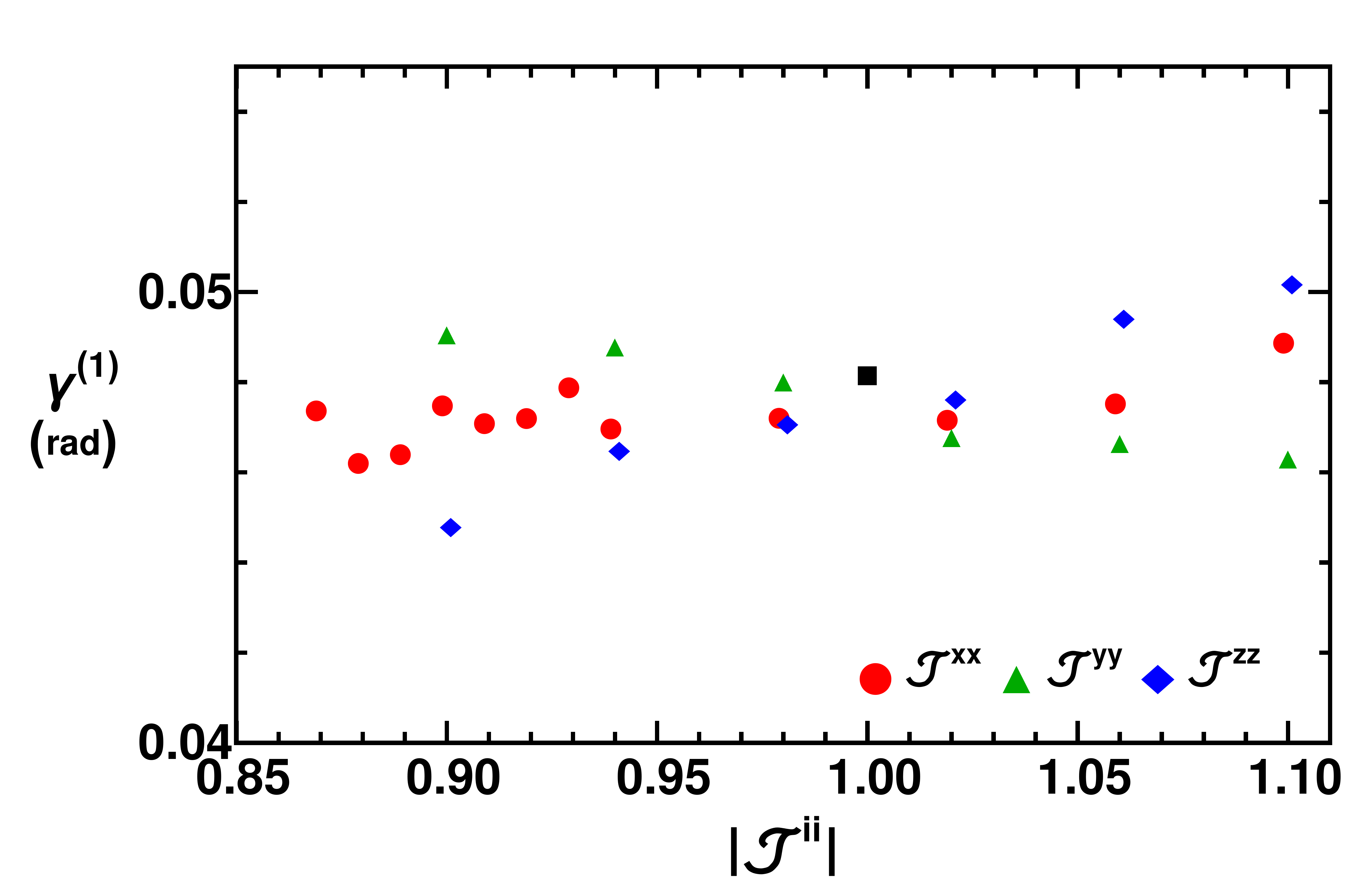}
    \end{subfigure}\\
    \begin{subfigure}[c]{0.45 \textwidth}
        \includegraphics[width=\textwidth,keepaspectratio]{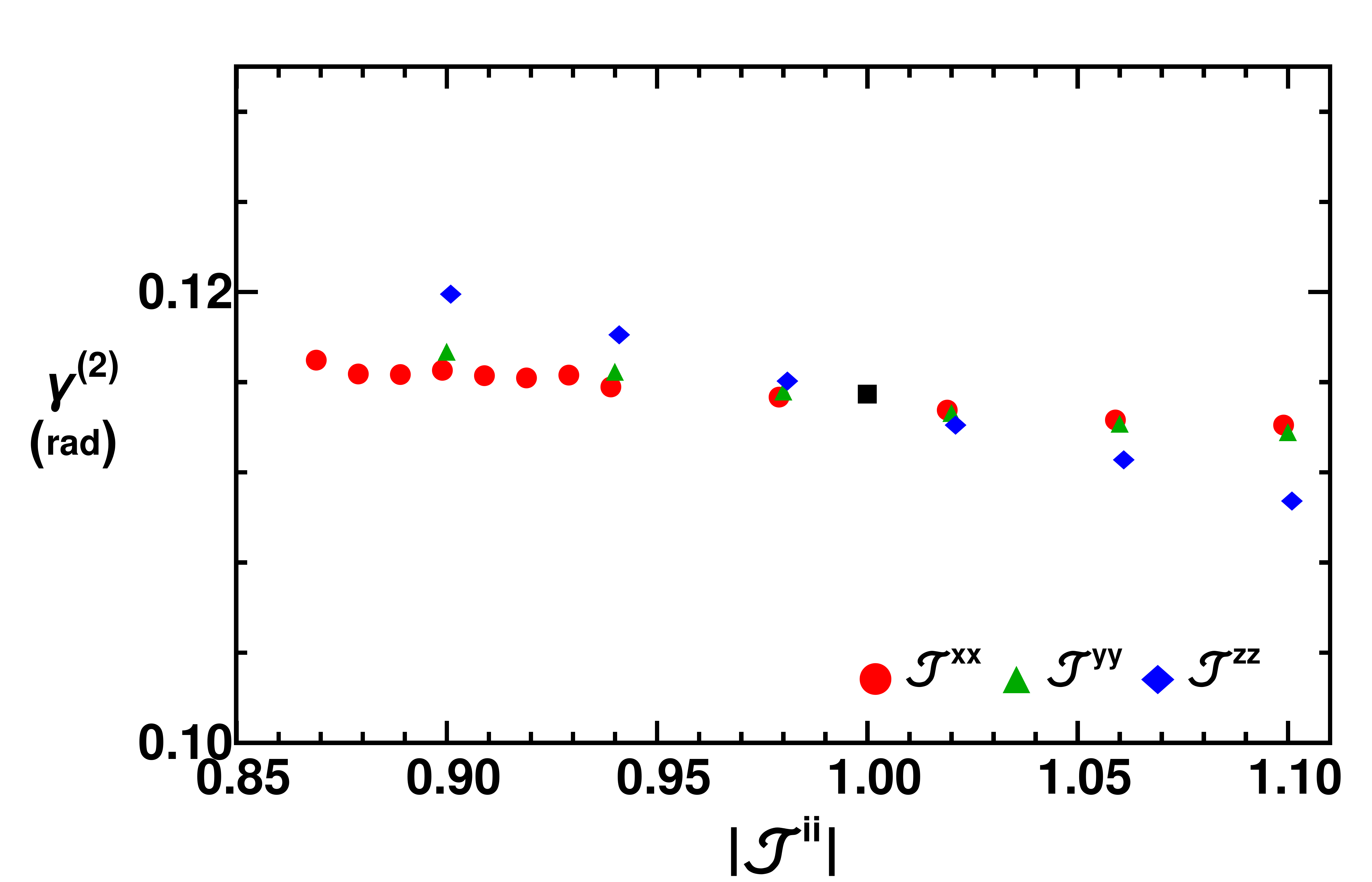}
    \end{subfigure}
    \caption[]{The out-of-plane angles 
    $\gamma^{(1)}$ and $\gamma^{(2)}$ as a function 
    of individual symmetric (Heisenberg-like) coupling constants. The black square represents the $|D| = 0.50$ result. 
  }
    \label{fig:Jii}
\end{figure}

Fig.\ \ref{fig:Jii} shows the dependence of the out-of-plane angles
 $\gamma^{(1)}$ and $\gamma^{(2)}$ on the three symmetric exchange
parameters ${\cal J}^{xx}$, ${\cal J}^{yy}$ and ${\cal J}^{zz}$.  For both angles we note the strongest dependence on ${\cal J}^{zz}$, while 
${\cal J}^{xx}$ yields nearly constant results. 
For $\gamma^{(1)}$, the slopes of the lines for ${\cal J}^{zz}$ and ${\cal J}^{yy}$ have  the opposite sign.  

Fig.\ 7 shows the dependence of $k$, $\phi$ and 
$\gamma^{(2)}$ on the anti-symmetric exchange
constants ${\cal J}_a^{ij}$. The wavenumber results are similar to what are shown
shown in Fig.~1, except for the variation of ${\cal J}_a^{zx}$  which has a linear dependence of the opposite sign. The plots for $\phi$ and
$\gamma_2^{(2)}$ also display linear dependence similar to what is shown  in Figs.\ 2 and 3, except for the 
parameter ${\cal J}_a^{xy}$, where the dependence is almost flat. It is clear from the plots that all three measurable quantities are sensitive to the tuning of the three
anti-symmetric exchange constants, especially the phase difference $\phi$, which varies by as much as 50\% in the (limited) range shown in the plot.

\begin{figure}[H]
    \begin{subfigure}[c]{0.45 \textwidth}
        \includegraphics[width=\textwidth,keepaspectratio]{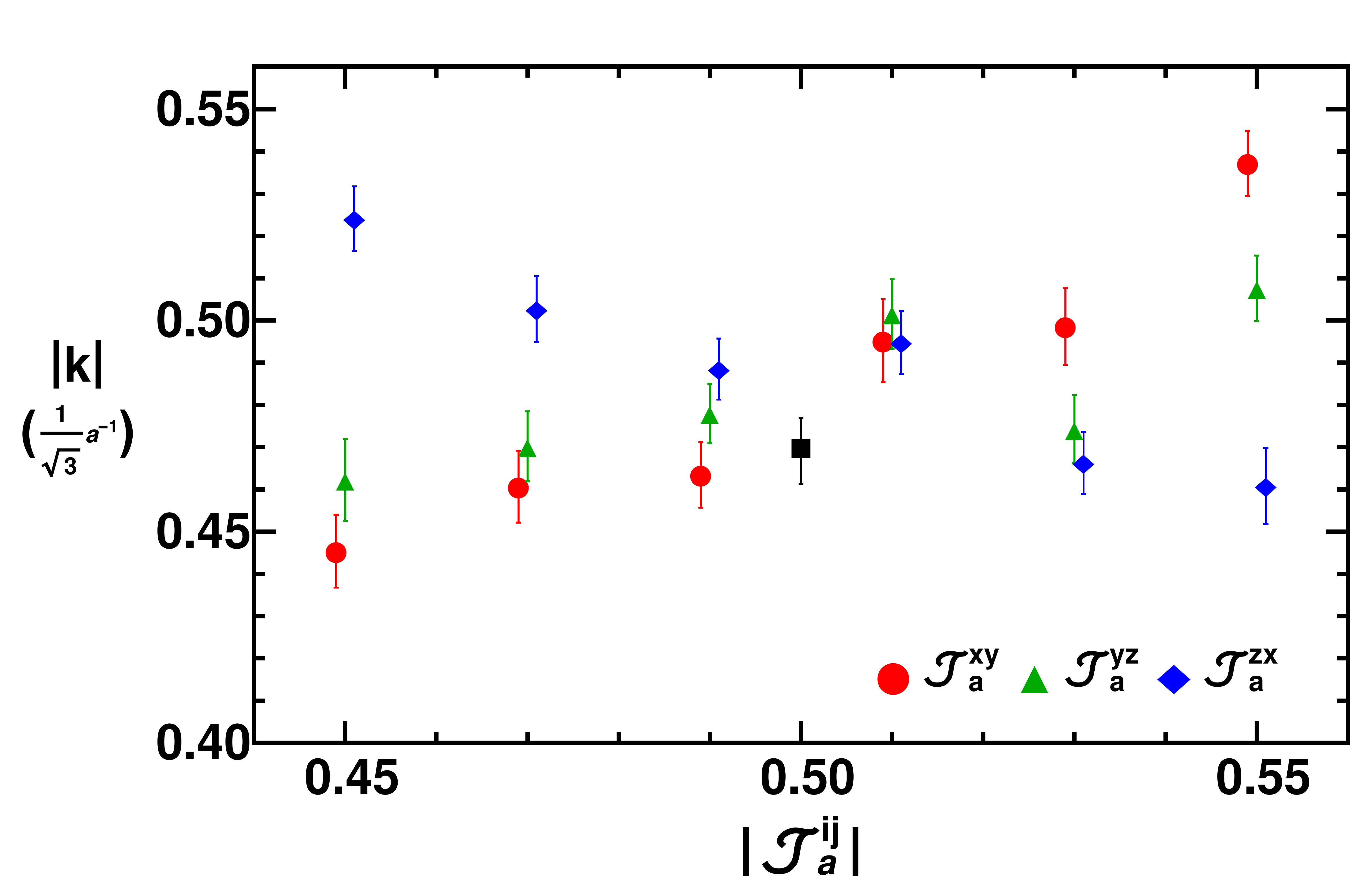}
    \end{subfigure}
    \begin{subfigure}[c]{0.45 \textwidth}
        \includegraphics[width=\textwidth,keepaspectratio]{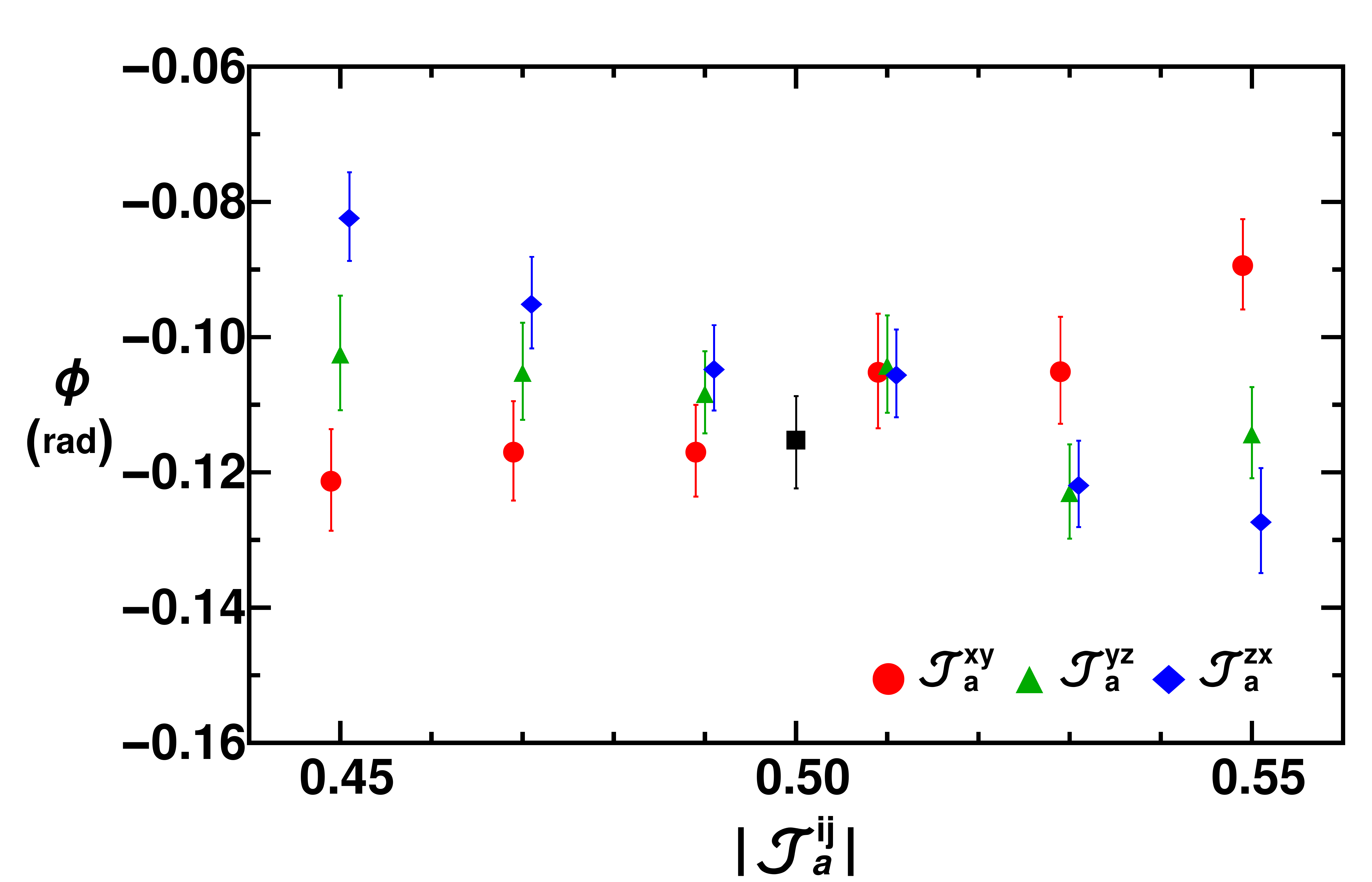}
    \end{subfigure} \\
    \begin{subfigure}[c]{0.45 \textwidth}
        \includegraphics[width=\textwidth,keepaspectratio]{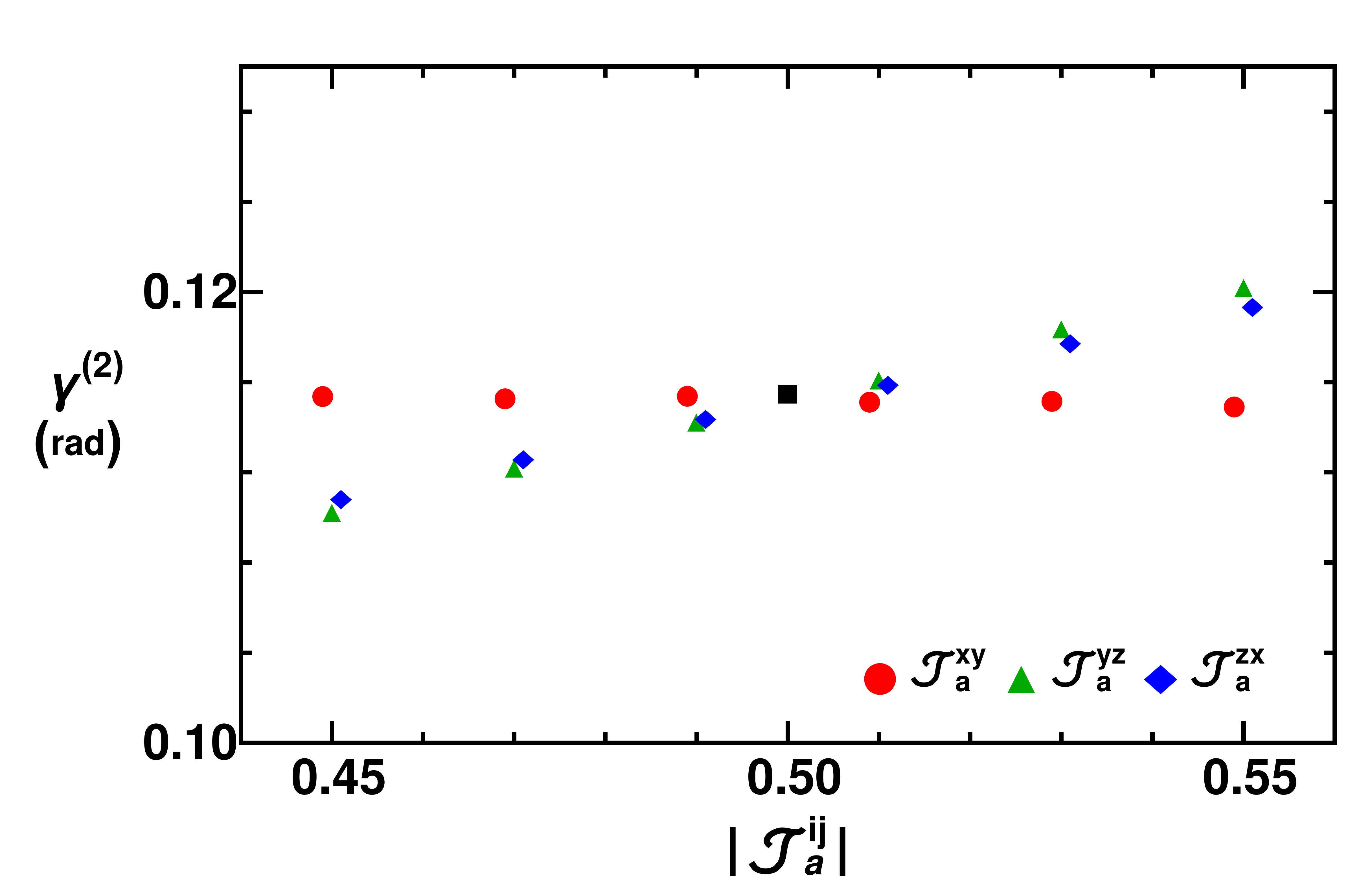}
    \end{subfigure}
    \caption[]{Wavenumber $k$, angle $\phi$, and out-of-plane angle $\gamma^{(2)}$ as functions of the antisymmetric coupling constants ${\cal J}_a^{ij}$. The black square represents the $|D| = 0.50$ result. In the bottommost plot the error bars are too small to be seen.  
    } 
\end{figure}

The independent variation of the anti-symmetric interaction constants was analyzed by 
 Chizhikov and Dmitrienko \cite{ChizhikovDmitrienko_PhysRevB.85.2012}, who used  the notation
 $(D_x,D_y,D_z) \equiv ({\cal J}^{zx}_a, {\cal J}^{xy}_a, {\cal J}^{yz}_a)$. 
 They found:
 \begin{eqnarray}\label{eq:DMIkRelate}
    k &= & \frac{2(D_{x} - 2 D_{y} - D_{z})}{3 J}
    \\
\label{eq:PhiDDependence}
    \phi &\propto &\frac{D_{x} + D_{z}}{J}
\\
\label{eq:PredCantingMag}
    \gamma &\propto &-\frac{(D_{x} + D_{z})}{J}
\end{eqnarray}
in rough agreement with the results shown in Fig.\ 7. In particular, these results predict the flatness of the plots for $\phi$ and $\gamma^{(2)}$ with respect to
the parameter ${\cal J}_a^{xy}$,
as well as the relative sizes and signs of the slopes of the plots for $k$.

 It is worth noting that the other six independent interaction terms also appear in 
 Ref.\ \cite{ChizhikovDmitrienko_PhysRevB.85.2012} with associated interaction constants. First there is a small 
 correction to the Heisenberg interaction constant $-J
 \rightarrow -J + \frac{D^2}{12 J}$
 with separate, additional contributions to each of the components in 
 $({\cal J}^{xx}, {\cal J}^{yy}, {\cal J}^{zz})$:
 $\frac{-1}{6J}(2D_z^2 - D_x^2 - D_y^2, 2D_x^2-D_y^2-D_z^2,
 2D_y^2- D_x^2 - D_z^2)$. The other symmetric interaction constants also appear as
  $({\cal J}^{yz}_s, {\cal J}^{zx}_s, {\cal J}^{xy}_s) \equiv \frac{-1}{2J}(D_xD_y, D_yD_z, D_zD_x)$.  
  However the analysis
  in Ref.\ \cite{ChizhikovDmitrienko_PhysRevB.85.2012} does not consider these contributions.

\subsection{Applied Field}\label{ssec:Results:AppField}

We also performed numerical simulations of the model under an applied magnetic field, where the field is scaled according to Eq.\ 3.  Fig.\ 8 shows the out-of-plane canting of the spins
as a function of an applied field in the 
$[111]$ direction.  As expected,
with increasing field,  canting toward the field direction increases as $\arcsin (B)$ and there is little difference between $\gamma^{(1)}$ and $\gamma^{(2)}$. Extrapolating the fit, complete alignment of the moments with the field occurs when $H \sim 1$.

\begin{figure}[H]
    \begin{subfigure}[c]{0.45 \textwidth}
        \includegraphics[width=\textwidth,keepaspectratio]{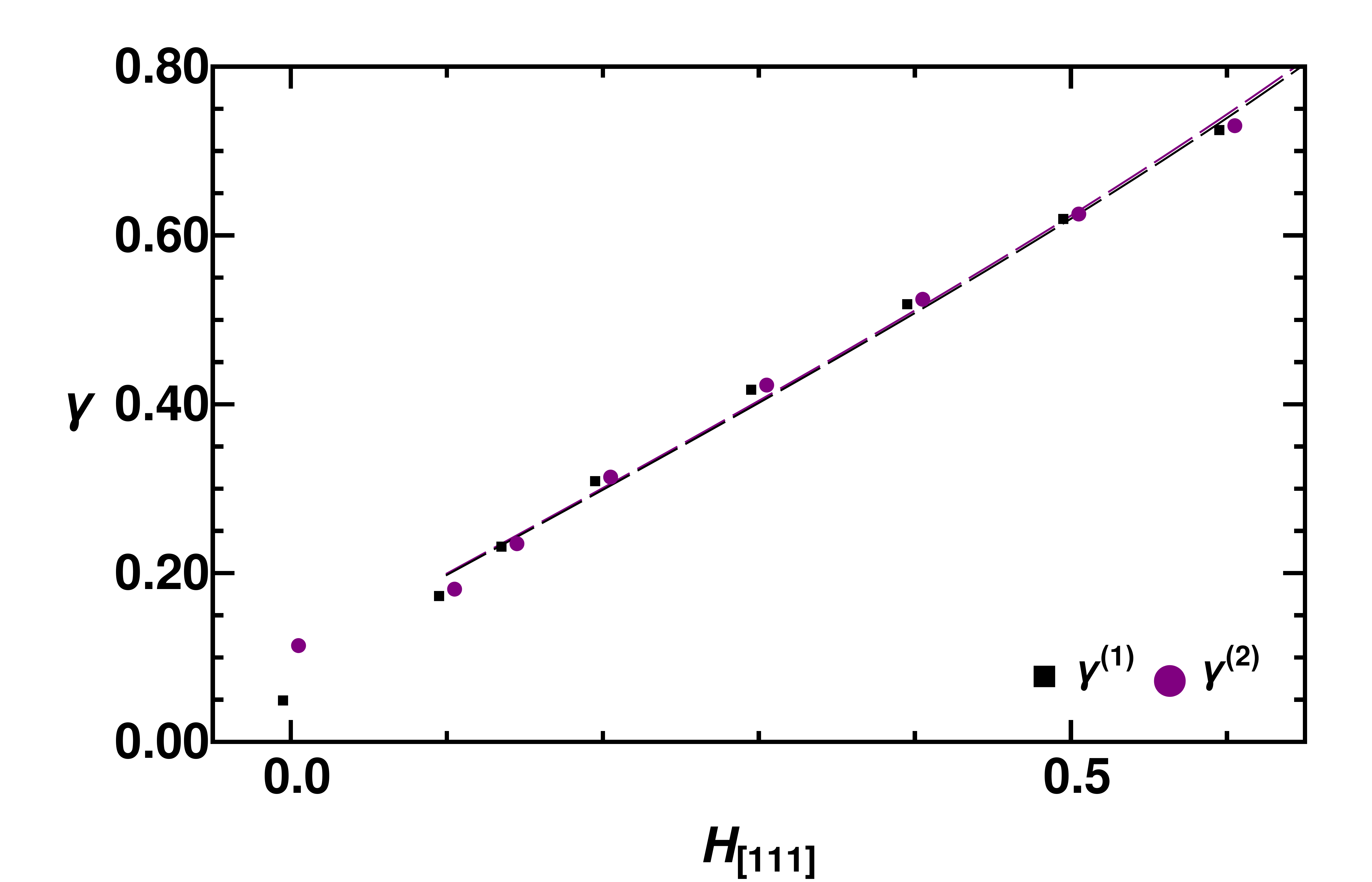}
    \end{subfigure}
    
    \caption[]{The out-of-plane canting of individual sublattices as a function of  applied field. The lines are fits to the function $c + \arcsin(a B)$. 
    }
   
\end{figure}

Fig.\ 9 shows how the components
of the order parameter evolve
as a function of an applied field in the
$[111]$ direction.  The $F_1$ components increase with increasing field, until they become the dominant contribution, in rough correspondence with the out-of-plane canting shown in Fig.\ 8. 
\begin{figure}[H]
    \centering
    \includegraphics[width = 0.45 \textwidth]{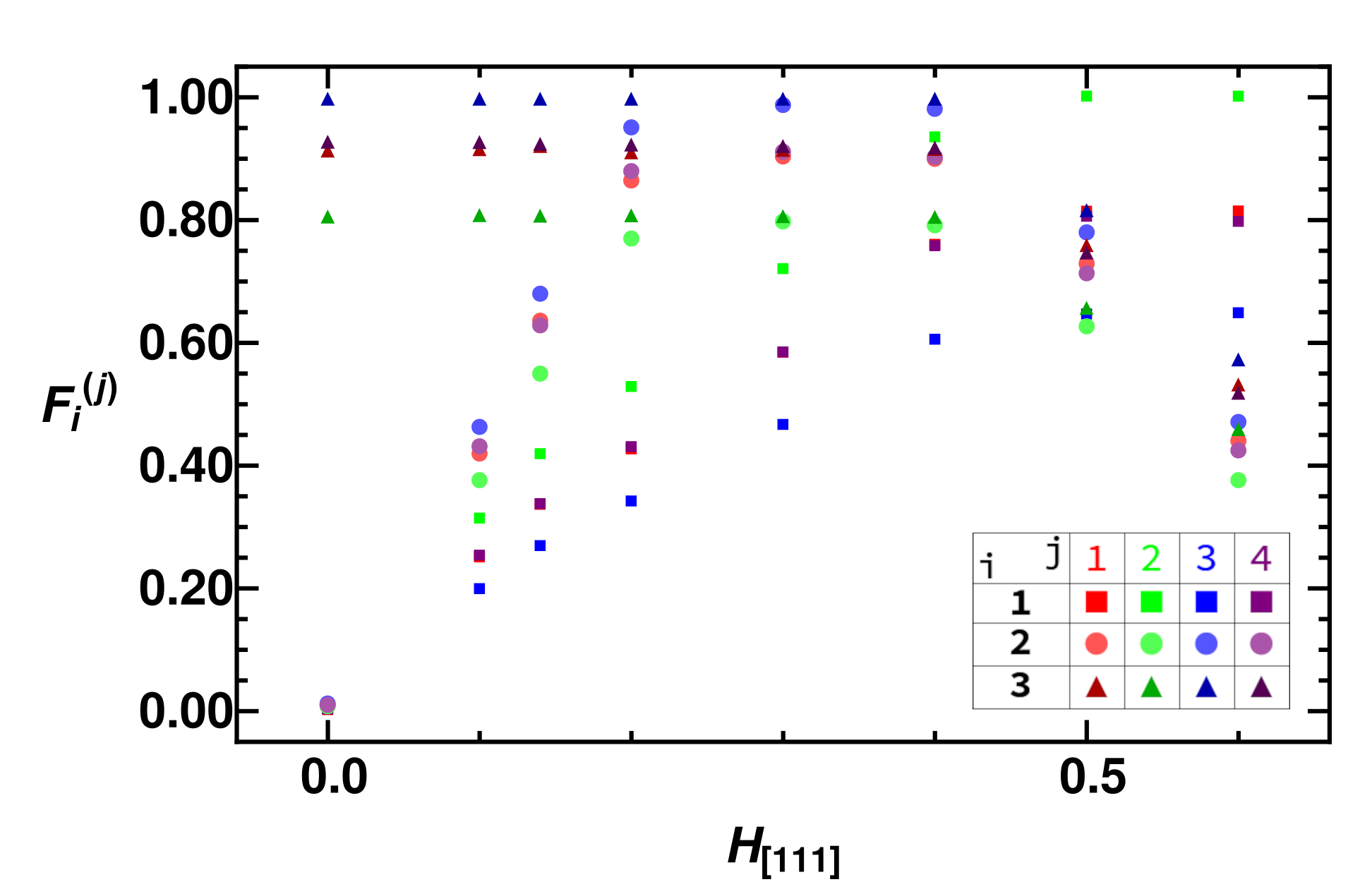}
    \caption{Magnetic order parameters vs applied field in the $[111]$ direction. The magnitudes are normalized within a single value of $H$ such that  the maximum is always $1.0$. 
    }
\end{figure}

\section{Discussion \& Conclusion}\label{sec:Discuss}

The computational results of the simplified two-parameter model discussed in Section IV A provide a qualitative description of the helical 
magnet phase of MnSi, but fail to describe it in detail: 
 the measured values of $k$ and 
$\phi$ cannot be consistently explained within a two-parameter model of 
symmetric and anti-symmetric interactions.  Our numerical simulations using the full model, presented in Section IV B, demonstrate
that an appropriate tuning of the parameters of the more general model could reproduce those experimentally measured values. 
Also, in our simulations, we have measured other features of the spin configuration in the helical phase that are potentially experimentally accessible - the out-of-plane angles $\gamma^{(1)}$ and $\gamma^{(2)}$. Our results are in good agreement with the free energy analysis of
Ref.\ \cite{ChizhikovDmitrienko_PhysRevB.85.2012}. Furthermore, an additional measurement, 
the canting of individual spins towards their $\tau_i$ axes (Eq.\ 21) could also be obtained from our results. 

We have also shown - at least within the limits of our finite-sized simulation - the extent to which the helical phase is contained within a single kind of order parameter (the $F_3$ OP), and the relative weight of the four separate contributions to $F_3$. Their evolution under an applied field was also presented.  These details can provide a different point of comparison for measurements on the helical phase.

To summarize, we have considered a general, symmetry-allowed model with eleven free parameters -  nine interaction constants and two same-site anisotropy constants - to describe the helical magnet phase in MnSi-like crystals. 
Experimental observations greatly constrain the parameter space of this model, as shown in the earliest theoretical studies, which derived the relationship between symmetric and anti-symmetric (DM) interaction constants and the spiral wavelength. Recent experiments have uncovered details concerning the orientation of individual magnetic moments in the magnetic spirals; those findings can be explained by the more detailed model we have considered here.

\begin{acknowledgments}
 This work was supported by the Natural Sciences and Engineering Research Council of Canada (RGPIN-05615-2020).
	\end{acknowledgments}
	\appendix

\bibliography{citations}

\end{document}